\documentstyle[12pt]{article}

\def\la{~\mbox{\raisebox{-.6ex}{$\stackrel{<}{\sim}$}}~}
\def\ga{~\mbox{\raisebox{-.6ex}{$\stackrel{>}{\sim}$}}~}

\begin{document}

\thispagestyle{empty}
\rightline{{\tt hep-th/0404005}}

\vskip 1cm \centerline{ \Large \bf Moving sources in a ghost condensate}

\vskip .2cm

\vskip 1.2cm

\centerline{ \bf Marco Peloso $^a$ and Lorenzo Sorbo $^b$}
\vskip 10mm \centerline{ \it $^a$ School of Physics and Astronomy,
University of Minnesota, Minneapolis, MN 55455, USA}
\vskip 3mm 
\centerline{\it $^b$ Department of Physics, University of California, Davis,
CA 95616, USA}
\vskip 1.2cm

\centerline{\tt  peloso@physics.umn.edu, sorbo@physics.ucdavis.edu}

\vskip 1.2cm

\begin{quote}

Ghost condensation has been recently proposed as a mechanism inducing the spontaneous breaking of Lorentz symmetry.  Corrections to the Newton potential generated by a static source have been computed: they yield a limit $M \la 10 \, {\rm MeV}$ on the symmetry breaking scale, and -  if the limit is saturated - they are maximal at a distance $L \sim 1000 \, {\rm km}$ from the source. However, these corrections propagate at a tiny velocity, $v_{\rm s} \sim 10^{-12}\,$m/s, many orders of magnitude smaller than the velocity of any plausible source. We compute the gravitational potential taking the motion of the source into account:  the standard Newton law is recovered in this case, with negligible corrections for any distance from the source up to astrophysical scales. Still, the vacuum of the theory is unstable, and requiring stability over the lifetime of the Universe imposes a limit on $M$ which is not too  far from the one given above. In the absence of a direct coupling of the ghost to matter, signatures of this model will have to be searched in the form of exotic astrophysical events.

\end{quote}

\newpage
\setcounter{footnote}{0}

\section{Introduction}

The whole interpretation of the currently available cosmological data strongly depends upon the hypothesis that the behavior of field theories -- and, in particular, of gravity -- at cosmological distances is the same one that we observe at local scales. It is hard to check the validity of such a theoretical assumptions by means of sufficiently prior-independent observations. It is therefore important to explore the possibility to modify gravity at large distances {\it in a theoretically consistent way}. Such goal is far from being trivially achievable. The simplest way to modify gravity in the infrared probably resides in the introduction of a tiny mass term for the graviton.  The introduction of a hard mass term, however, leads to a series of problems, which show some similarity with the ones encountered when introducing a mass term for a spin-one gauge boson. Consistency of the theory requires the mass term to be of the Fierz-Pauli type~\cite{fp}, and the propagator for the graviton is affected by the van Dam-Veltman-Zakharov discontinuity~\cite{dvz}, related to the different number of polarizations between the massive and the massless case. In addition, due to the specific form of the kinetic term for the longitudinal component, the perturbative description of gravity breaks down at macroscopic lengths, that are probed in everyday life~\cite{vai}. These properties can be nicely understood with the technique of~\cite{ags}, where the gravitational counterpart of the Goldstone description of massive gauge theories was constructed.

The recent years have witnessed a wide debate about the modifications of gravity at large scales in higher dimensional models, in which the four dimensional graviton emerges as a resonance, and the higher--dimensional theory shows up at large distances as well as at short ones~\cite{extra}. In these scenarios, four-dimensional covariance is preserved both for the theory and for its vacuum state. If we instead restrict the attention to four-dimensional scenarios, and in view of the analogies of General Relativity with flat space gauge theories, it is natural to expect that a Higgs phenomenon could be a simple way to give consistently a mass to the graviton. Indeed, spontaneous breaking of Lorentz symmetry has been studied by several authors (see, for instance~\cite{lsb}) in the past years, with a main interest in the quantum gravitational origin of this phenomenon. Recently, a further progress in this direction has been made in~\cite{ghost}, where the spontaneous breaking of (part of) the Lorentz group is achieved by giving a (timelike) expectation value to the gradient of a scalar field $\phi\,$.~\footnote{The work~\cite{ghost} was accompanied by~\cite{paolo}, which studied inflation in this theory. The primordial perturbations generated during inflation were found to have distinctive signatures with respect to the standard results of slow roll inflation.} When close to its Lorentz invariant unstable equilibrium state, this field behaves as a ghost. For this reason, this mechanism has been named {\it ghost condensation}. The scale of the transition to the broken phase is set by a fundamental dimensionful parameter $M\,$. At energy larger than $M$, the theory needs a UV completion that should presumably describe the emergence of a symmetric phase. At scales below $M$, Lorentz symmetry is broken.  The proposal of~\cite{ghost} has the virtue of being clearly treatable by perturbative methods all the way up to the symmetry breaking scale.~\footnote{This is due to the fact that $\phi$ has a well behaving kinetic term in its ground state in the absence of gravity.} The breaking of Lorentz symmetry is associated to some very unusual features, the most striking of which is probably the nonrelativistic dispersion relation $\omega^2=k^4/M^2$ for the fluctuations of $\phi\,$.

When the ghost is coupled to gravity, its dispersion relation is modified: the system develops a (Jeans-like) instability in the IR. Such effect can be used to set model independent\footnote{Direct couplings of the ghost to matter are expected to emerge at least through loop effects. Such couplings have to be strongly suppressed in order not lead to disagreement with observation, and depend on the details of the theory. In order to stick only to the minimal features of the scenario, we will not consider these couplings here.} bounds on the scale $M$, that fixes the time and the length scales of the gravitational instability of the model. Indeed, in the Newtonian limit $\omega^2\ll k^2$, the dispersion relation of the scalar degree of freedom of the system reads
\begin{equation}
\omega^2=\frac{k^4}{M^2}-\frac{M^2}{M_p^2}\,k^2\,\,.
\end{equation}
For sufficiently small $k$, $\omega$ turns out to be imaginary. The corresponding instability is maximal for wavenumbers $k\sim m\equiv M^2/M_p$ and frequencies $\left\vert \omega\right\vert\sim \Gamma\equiv M^3/M_p^2$. Going to real space, the instability is thus expected to develop on timescales $\tau \sim \Gamma^{-1}$ and on lengthscales $L \sim m^{-1}\,$. Notice that, due to the breaking of Lorentz invariance, the typical space-- and time--scales for the instability can be very different: unless $M$ is close to the Planck scale, $\tau$ will be much larger than $L\,$.

For the reasons we mentioned, the proposal of~\cite{ghost} is very interesting, and it is worth to subject it to close scrutiny. In order to set constraints on the parameter space of the model, it is crucial to establish the observables over which the instability will leave its strongest imprint. A natural candidate is the gravitational force between two masses. Indeed, the computation~\cite{ghost} of the gravitational potential generated by a static source shows, in addition to the standard Newton potential, a correction which is growing with time. In agreement with the above discussion, the growth occurs on a timescale $\tau\,$. In order to make sure that the new term has not yet grown to an observable size, one can conservatively require this time to be greater than the age of the Universe\footnote{In this analysis, the effects related to the expansion of the Universe are neglected. Due to the magnitude of the scales in consideration, we do not expect these effects to change significantly the present discussion.}. This results in a rather stringent upper bound on the scale of the condensate, $M \la 10 \,{\rm MeV}\,$. However, after a time $\tau$, the exponential growth will have occurred only at a distance $r \la L$ from the source~\cite{ghost}. For such a value of $M\,$, one finds $L \simeq 1000 \, {\rm km}\,$. At these scales, an oscillating modulation of the newtonian potential around a static source would be a distinctive signature of the mechanism. In the present note we will discuss how this picture is changed in the case in which the sources are in motion.

The mechanism of~\cite{ghost} breaks only a part of Lorentz symmetry. For a time-like $\langle \partial_\mu \phi \rangle \,$, we can choose a frame in which only $\langle \partial_t \phi \rangle$ is nonvanishing. We call this coordinate system the rest frame of the ghost. The presence of a privileged system allows to speak about absolute motion: $\phi-$mediated interactions between particles will depend on their velocity (even if constant) with respect to the ghost rest-frame. The standard newtonian interaction propagates instantaneously (the whole computation is performed in the nonrelativistic regime). However, the unstable modes of the ghost condensate propagate with an extremely low velocity $v_{\rm s}$. As we remarked, for $M\sim 10 \, {\rm MeV}\,$, it takes a time comparable to the age of the Universe for the modified interaction to propagate at a distance $\sim 1000 \, {\rm km}$ from the source. This gives $v_{\rm s} \sim M/M_p \sim 10^{-12}\,$m/s. Even if we do not know which the rest-frame of the condensate is, the typical velocity of celestial bodies is of the order of $10-100\,$km$/$s. This is the case, for instance, both for the motion of the Earth around the Sun, and of the Earth in the frame in which the Cosmic Microwave Background (CMB) radiation shows no dipole. As a consequence, any realistic source will be moving with respect to the ghost rest frame with a velocity that exceeds $v_{\rm s}$ by many orders of magnitude.

We compute the gravitational potential for a source moving with velocity $v \gg v_{\rm s}$ in the next Section. As in~\cite{ghost}, we assume that the source has been created (for instance, by gravitational collapse) at some finite time $t=0\,$. We then specify the general expression to two simplifying situations. Eq.~(\ref{nearsource}) describes the potential measured by a (late time) observer at rest with respect to the source. This observer {\it does not} see an exponentially growing correction to the Newton potential. Both the source and the observer are moving too fast with respect to the rest frame of the condensate, and the $\phi-$mediated interactions will not be observable. 

The situation is different if the gravitational potential is computed at late times in the vicinity of the place where the source was created. In this case, the nonstandard term is indeed growing exponentially. Perhaps most surprisingly, the exponential growth is taking place even if the source has in the meantime moved to very faraway distances! This shows that the growth of the potential is not really  due to the presence of a nearby source at all times, but rather to the instability of the theory. The source is simply triggering the initial instability, and a time $t \ga \tau\,$ is then needed for the instability to grow at exponentially large values. The massive source is required to start this growth, since the calculation performed here is classical. Most probably, quantum effects will also trigger the instability, even in the case in which classical sources were absent.

Although we do not expect signatures in the form of a modified Newton law, the instability of the vacuum will lead to potentially observable effects, if the scale of the instability is not too far from to the present age of the Universe (that is, $M\sim 10$~MeV). Unstable regions will presumably collapse to a strong gravity regime, and one may expect formation of compact objects, possibly leading to exotic astrophysical events. Such structures could be located in regions which are not correlated to any visible matter (since the source has in the meantime moved away) and hence they would be hard to explain in more conventional scenarios.

\section{Modification of the gravitational potential}

As in~\cite{ghost}, we compute the gravitational potential generated by a source of mass $\mu$
``nucleated'' at a given time $t_{\rm in}=0\,$ at the position ${\bf r} = {\bf 0}\,$. We assume that, after it is nucleated, the source moves with a constant velocity ${\bf v}$ in the rest-frame of the condensate. This extends the computation of~\cite{ghost} performed for $v = 0\,$.  The gravitational potential generated by the source is, at a generic position $r$ and at the time $t > 0\,$,
\begin{eqnarray}
V &\equiv& V_{\rm n} + \Delta \, V =  \frac{\mu \, G}{4 \, \pi^3} \, \int d^3 {\bf r}' \, d t' \, d^3 {\bf k} \, d \omega
\times \nonumber\\
&&\times  {\rm e}^{\,-i\,\omega \left( t - t' \right)+ i \, {\bf k} \cdot \left( {\bf r} - {\bf r}' \right)} \, \theta \left( t' \right) \, \delta^3 \left( {\bf r}' - {\bf v} \, t' \right) \, \Pi \,\,,
\end{eqnarray}
where $G = 1/ \left( 8 \, \pi M_p^2 \right)$ is the Newton constant. $\Pi$ is the propagator (in Fourier space) of the scalar perturbations of the metric~\cite{ghost}
\begin{equation}
\Pi \equiv - \frac{1}{k^2} + \frac{\alpha^2 \, m^2 / M^2}{\omega^2 + \frac{\alpha^2 \, m^4}{M^2} \, \left[ \left( \frac{k}{m} \right)^2 - \left( \frac{k}{m} \right)^4 \right]} \,\,,
\label{propago}
\end{equation}
which is valid in the nonrelativistic limit $\omega \ll k\,$. The parameter $M$ is the scale of the ghost condensate, while 
\begin{equation}
m \equiv M^2 / \left( \sqrt{2} \, M_p \right) \simeq 3 \cdot 10^{-19} \, {\rm GeV} \left( \frac{M}{\rm GeV} \right)^2 \,\,.
\end{equation}
Finally, $\alpha$ is model dependent parameter of order one (in the following we simply set $\alpha = 1\,$).

Let us define ${\bf {\tilde r}} \equiv {\bf r} - {\bf v} \, t$ to be the position of a given point in the rest frame of the source. The first term in~(\ref{propago}) gives rise to the standard newtonian potential $V_{\rm n} = - G \, \mu / \vert {\bf {\tilde r}}  \vert \,$, dependent on the (instantaneous~\footnote{In the nonrelativistic approximation that we are here considering, the Newtonian interaction propagates instantaneously.}) distance from the source only. To compute the second term, we first perform the trivial $d^3 {\bf r'}\,$ integration,
\begin{eqnarray}
\Delta \, V = \frac{G \, m \, \mu}{4 \, \pi^3} \int_0^\infty d T' \int d^3 {\bf u} \int d W {\rm e}^{-i W \left( T - T' \right) + i {\bf u} \cdot {\bf R} - i {\bf u} \cdot {\bf {\cal V}} T'} \frac{1}{W^2+u^2-u^4}
\end{eqnarray}
where we have introduced the following adimensional quantities
\begin{eqnarray}
&& u \equiv k / m \quad,\quad W \equiv M \, w / m^2 \,\,, \nonumber\\
&& R \equiv m \, r = 1.5 \left( \frac{M}{\rm GeV} \right)^2 \left( \frac{r}{\rm km} \right)  \,\,, \nonumber\\
&& {\cal V} \equiv \frac{M \, v}{m} = 3.3 \cdot 10^{15} \left( \frac{\rm GeV}{M} \right) \, \left( \frac{v}{10^{-3} \, c} \right) \,\,, \nonumber\\
&& T \equiv \frac{M^3}{2 \, M_p^2} \, t = 6.2 \cdot 10^{4} \left( \frac{M}{\rm GeV} \right)^3 \, \left( \frac{t}{t_0} \right) \,\,, \nonumber\\
&& {\tilde R} \equiv m {\tilde r} = m \left( r - v \, t \right) = R - {\cal V} \, T \,\,,
\label{para}
\end{eqnarray}
where $c$ is the speed of light, while $t_0 \simeq 15$ billion years is the age of the Universe.

To proceed further, we integrate over complex frequencies $\omega$ with a given prescription for the contour. This term is analogous to the propagator of a tachyonic field: for small momenta ($k < m\,$, in our case) the frequency has imaginary poles, which  are related to the tachyonic instability of the vacuum; high momenta $k > m$ balance this effect, and these modes are stable. Causality arguments prescribe the use of the retarded propagator ($\omega \rightarrow \omega + i \, \epsilon$) for these modes. For low momentum modes the choice of the contour is instead more ambiguous. We deform the contour so that it never crosses the poles, when - as $k$ decreases - they go from real to imaginary. Mathematically, the case $k < m$ becomes the analytic continuation of $k > m\,$. Physically, different contours correspond to different initial conditions, and the one we choose gives $\Delta V = 0$ at the initial time $t=0\,$. Taking this into account, we find
\begin{eqnarray}
&&\Delta V = - \frac{G \, m \, \mu}{2 \, \pi^2} \times \nonumber\\
&&\!\!\!\!\!\!\left\{ \: \int_{u \leq 1} d^3 {\bf u} \frac{{\rm e}^{\, i \, {\bf u} \cdot {\bf R}}}{u \sqrt{1-u^2}}
\frac{u \sqrt{1-u^2} \, {\cal Q}_1 + i \, {\bf u} \cdot {\bf {\cal V}}  \, {\rm sinh}  \left( u \sqrt{1-u^2} \, T \right)} {u^2 \left( 1 - u^2 \right) + \left( {\bf u} \cdot {\bf {\cal V}} + i \, \epsilon \right)^2} \,\,, \right. \nonumber\\ 
&& \left. \int_{u \geq 1} d^3 {\bf u} \frac{{\rm e}^{\, i \, {\bf u} \cdot {\bf R}}}{u \sqrt{u^2-1}}
\frac{u \sqrt{u^2-1} \, {\cal Q}_2 + i \, {\bf u} \cdot {\bf {\cal V}} \, {\rm sin}  \left( u \sqrt{u^2-1} \, T
\right)} {u^2 \left( u^2 - 1 \right) - \left( {\bf u} \cdot {\bf {\cal V}} + i \, \epsilon \right)^2} \right\} \nonumber\\
\label{preangular}
\end{eqnarray}
where we have defined
\begin{eqnarray}
{\cal Q}_1 &\equiv& {\rm e}^{-i \, {\bf u} \cdot {\bf {\cal V}} T} - {\rm cosh } \left( u \sqrt{1-u^2} \, T \right) \,\,, \nonumber\\
{\cal Q}_2 &\equiv& {\rm e}^{-i \, {\bf u} \cdot {\bf {\cal V}} T} - {\rm cos } \left( u \sqrt{u^2-1} \, T \right) \,\,,
\end{eqnarray}
and where $u \equiv \vert {\bf u} \vert \,$. 

Already at this stage we can appreciate the effect of the velocity of the source. In the rescaled units, the speed of propagation of $\Delta V$ (see the previous Section) corresponds to ${\cal V}_{\rm s} \simeq 1 \,$. Sources which move much faster than $v_{\rm s}$ will have ${\cal V} \gg 1\,$. 
In this case, we then expect the ${\cal V}-$dependent terms to be important in~(\ref{preangular}).

The angular integration can be performed exactly when ${\bf {\tilde r}}$ and ${\bf v}$ are parallel, that is when the potential is computed along the line of motion of the source. The following explicit computation is restricted to this case; however, the arguments presented above and our conclusions do not depend on this assumption. The angular integrals then give

\begin{eqnarray}
\Delta V &=& \frac{G \, m \, \mu}{\pi \, {\cal V}} \left\{ \, \int_0^1 \frac{d u}{\sqrt{1-u^2}} \left[ {\rm cosh } \left( b \, u \, {\tilde R} \right) {\cal A} + {\rm sinh } \left( b \, u \, {\tilde R}  \right) {\cal B} \right] + \right. \nonumber\\
&& \left. \quad\quad\quad + \int_1^\infty \frac{d u}{\sqrt{u^2-1}} \left[ {\rm cos } \left( b \, u \, {\tilde R} \right) {\cal C} + {\rm sin } \left( b \,u \, {\tilde R} \right) {\cal D} \right] \; \right\} \,\,, \nonumber\\
&&\nonumber\\
{\cal A} &\equiv& \left[ \pi \, \sigma \left( {\tilde R} \right) + i \, CI \left( u {\tilde R} \left( 1 + i \, b \right)
\right) - i \, CI \left( u {\tilde R} \left( 1 - i \, b \right) \right) \right] + \nonumber\\
&&- \left[ {\tilde R} \rightarrow R \right] \,\,, \nonumber\\
{\cal B} &\equiv& \left[ - SI \left( u {\tilde R} \left( 1 + i \, b \right) \right) - SI \left( u {\tilde R} \left( 1 - i \, b \right) \right) \right] - \left[ {\tilde R} \rightarrow R \right] \,\,, \nonumber\\
{\cal C} &\equiv& \left[ - i \, \pi \, \theta \left( b - 1 \right) \sigma \left( {\tilde R} \right) - CI \left( u {\tilde R} \left( 1 + b \right) + i \, \epsilon \right) + \right. \nonumber\\
&& \left. + CI \left( u {\tilde R} \left( 1 - b \right) + i \, \epsilon\right) \right] - \left[ {\tilde R} \rightarrow R \right] \,\,, \nonumber\\
{\cal D} &\equiv& \left[ - SI \left( u {\tilde R} \left( 1 + b \right) \right) - SI \left( u {\tilde R} \left( 1 - b \right) \right) \right] - \left[ {\tilde R} \rightarrow R \right] \,\,, \nonumber\\
b &\equiv& \frac{\sqrt{\vert 1 - u^2 \vert}}{\cal V} \,\,,
\label{pot}
\end{eqnarray}
where $\sigma \left( x \right)$ is the sign of $x\,$, while SI and CI denote, respectively, the sine and cosine integral, defined as~\cite{librosacro}
\begin{equation}
SI \left( x \right) \equiv - \int_x^\infty d t \: \frac{{\rm sin } \, t}{t}
\quad,\quad
CI \left( x \right) \equiv - \int_x^\infty d t \: \frac{{\rm cos } \, t}{t} \,\,.
\end{equation}

At the initial time, $R={\tilde R}$ and, as we noted, $\Delta V = 0\,$. In addition, one can verify that~(\ref{pot}) reproduces the analogous expression of~\cite{ghost} as the velocity ${\cal V}$ is sent to zero. The first line of~(\ref{pot}) describes the tachyonic modes, and we will focus only on the calculation of this piece (we denote it by $\Delta V_1$) in the remaining of the Section. The second line describes the stable modes, and indeed it does not contain any term which grows exponentially with time. For this reason, it is not responsible for the effect we are considering, and, as done in~\cite{ghost}, we can simply ignore it.  The expression~(\ref{pot}) is still exact~\footnote{More precisely, it holds in the nonrelativistic limit $\omega \ll k\,$. For $\Delta V_1\,$, the poles are at $\vert \omega \vert \sim \vert k \vert m/M\,$. Hence, the nonrelativistic limit is correct as long as $M \ll M_p\,$. }. To proceed further, we 
can approximate the integrand for ${\cal V} \gg 1$ (we recall that ${\cal V} = 1$ corresponds to $v = v_{\rm s}\,$).
\begin{eqnarray}
\Delta V_1 &\simeq& \frac{2 \, G \, m \, \mu}{\pi \, {\cal V}}
\int_0^1 \frac{d u}{\sqrt{1-u^2}} \, \times \nonumber\\
&& \Bigg\{ b \bigg[ \Big[ {\rm cos } \left( u \vert R \vert \right) + u \vert R \vert \,
SI \left( u \vert R \vert \right) \Big] 
\, {\rm cosh } \left( u \, b \left( R - {\tilde R} \right) \right) - {\rm cos } \left( u \vert {\tilde R} \vert \right) +   \nonumber\\
&& \;\;\;\;\;\;\; - u \vert {\tilde R} \vert \,
SI \left( u \vert {\tilde R} \vert \right) \bigg] 
- \sigma \left( R \right) \, SI \left( u \vert R \vert \right) \,
{\rm sinh } \left( u  \left( R - {\tilde R} \right) \, b \right) 
+ \nonumber\\
&& \;\;\; + \pi \, {\rm sinh } \left( u \, b \vert {\tilde R} \vert \right)
\sigma \left( {\tilde R} \right) \left[ \theta \left( - {\tilde R} \right) - \theta \left(
- R \right) \right] \Bigg\} \,\,.
\label{largev}
\end{eqnarray}
This result holds for $b \equiv \sqrt{1-u^2} / {\cal V} \ll 1\,$, but arbitrary $b\, \vert R \vert$ and $b \, \vert {\tilde R} \vert \,$. Much simpler expressions can be obtained in the limits of very small or large $R,\; {\tilde R}\,$. In particular, we can distinguish two relevant cases, which we have already discussed in the previous Section.

1) ${\cal V \, T} \gg 1 \;,\; \vert {\tilde R} \vert \ll {\cal V} \;,\; R \gg T \;,\; R \gg 1 \,$. In this case we find

\begin{eqnarray}
\Delta V_1 \simeq \left\{
\begin{array}{c}
- \, \frac{2\,G\,\mu\,m}{\pi\,{\cal V}^2} \quad\quad\quad\quad\quad\;,\;\;\quad \vert {\tilde R} \vert 
\ll 1 \,\,,\\ \\
\frac{G \, \mu \, m}{{\cal V}^2} \left[ {\tilde R} \, \theta\left( - {\tilde R} \right) - 2\,\frac{{\rm cos } {\tilde R}}{{\tilde R}^2} \right] \quad, \quad \vert {\tilde R} \vert \gg 1 \,\,.
\end{array}
\right.
\label{nearsource}
\end{eqnarray}
this computation applies for a late time observer at a fixed distance ${\tilde R}$ (taken to be small, on cosmological scales) from the source. As we explained in the previous Section, $\Delta V$ {\em does not} grow exponentially in this case. To clarify why this happens, let us compute in more details $\Delta V_1$ for ${\tilde R} =0\,$, that is at the exact location of the (moving) source. Starting from~(\ref{pot}), and expanding CI and SI for large argument, we get
\begin{equation}
\Delta V_1 \left( {\tilde R} = 0 \right) \! = \! \frac{2 \, G \, m \, \mu}{\pi \, {\cal V}^2} \!\! \int_0^1 \! d u \left[ 
\frac{{\rm cos } \left( u \, {\cal V} \, T \right) \, {\rm sinh } \left( u \sqrt{1-u^2} \, T \right)}{u \sqrt{1-u^2} T} - 1 \right] + {\rm O } \left( \frac{1}{{\cal V}^3} \right) .
\label{rt0}
\end{equation}
This expression contains the function ${\rm sinh } \left( u \sqrt{1-u^2} \, T \right) \,$, which is growing exponentially with time. However, it is multiplied by the much faster oscillating function ${\rm cos } \left( u \, {\cal V} \, T \right)\,$. Since ${\cal V} \gg 1\,$, positive and negative part of the integrand average to a very small value, which does not grow with time. This suppression effect can be explicitly seen to be at work already at the level of eq.~(\ref{pot}), unless $R$ is small (that is, close to the point where the source was nucleated, see below). The first term of~(\ref{rt0}) cannot be computed exactly. We obtain an approximate primitive by noticing that, for ${\cal V} \gg 1\,$,
\begin{equation}
\frac{{\rm cos } \left( u \, {\cal V} \, T \right) \, {\rm sinh } \left( u \sqrt{1-u^2} T \right)}{u \sqrt{1-u^2}}
\simeq \frac{d}{du} \left[ \frac{{\rm sin } \left( u \, {\cal V} \, T \right) \, {\rm sinh } \left( u \sqrt{1-u^2} T \right)}{{\cal V} \, T \, u \sqrt{1-u^2}} \right]
\label{primitiva}
\end{equation}
(in practice, we simply use the primitive of the most rapidly oscillating function; we have verified numerically that this approximation is accurate\footnote{Alternatively, one can proceed as in~\cite{ghost}, by substituting ${\rm sinh } \left( u \sqrt{1-u^2} T \right),$ with a gaussian, and by extending the integration from $0 < u < 1$ to $- \infty < u  < + \infty\,$.
This approximation is valid only in the limit ${\cal V}\ll 1$, and would give a result $\propto {\rm exp } \left( - {\cal V}^2 \, T / 8 \right)\,$.}). Inserting eq.~(\ref{primitiva}) into~(\ref{rt0}) we recover the first line of~(\ref{nearsource}).

The result~(\ref{nearsource}) has a few other features which is worth noting. First, it is time independent. This was expected, since the potential close to the source resents only negligibly of the previous history of the source , due to the fact that both the source and the observer are moving much faster than the speed at which the signal is propagating.~\footnote{Strictly speaking, this is not true in the transient $T \ll 1\,$ regime, since in this case one cannot neglect the fact that the source was nucleated at the finite time $T=0\,$.} Finally, we also note that, for $1 \ll \vert {\tilde R} \vert \ll {\cal V} \,$, the potential is greater at negative ${\tilde R}\,$. This is easily understood by remembering that negative ${\tilde R}$ describe points where the source has already passed, and where the instability has started to develop, although the growth is still in the linear regime. Positive ${\tilde R}$ are instead points at which the source has yet to come.

2) The second interesting regime for the computation of~(\ref{pot}) is $\vert R \vert \ll {\cal V} \;,\; \vert {\tilde R} \vert \simeq {\cal V} T \, \gg {\cal V} \,$. In this case we get

\begin{eqnarray}
\Delta V_1 \simeq \left\{
\begin{array}{c}
- \frac{G\,m\,\mu\,\sqrt{\pi}}{2\,{\cal V}\,\sqrt{T}}\,{\rm e }^{T/2}
\quad\quad\quad\;,\quad
\vert R \vert \ll 1\\ \\
- \frac{G\,m\,\mu\,\sqrt{\pi}}{2\,{\cal V}\,\sqrt{T}}\,{\rm e }^{T/2} \left[ 2 \,
\theta \left( R \right) +{\rm O}\left(\frac{1}{R}\right) \right] , \\ \\ \quad\quad\quad\quad 1 \ll \vert R \vert \ll T
\end{array} \right.
\label{farsource}
\end{eqnarray}
This computation applies for a late time observer which is at rest in the rest-frame of the condensate, and close to where the source was originally nucleated. In this case, $\Delta V$ is growing exponentially with time. The exponential growth is related to the instability of the vacuum triggered by the source when it was passing through these points, as discussed in the previous Section.~\footnote{The growth visible in~(\ref{farsource}) is analogous to the one computed in~\cite{ghost} for a static source, with the difference of the strong suppression factor $1/{\cal V}\,$.}

\section{Conclusions}

Effects associated to the breaking of Lorentz symmetry will in general depend
upon the motion of the observer with respect to the preferred frame of the
theory. In particular, a phenomenologically very interesting aspect of ghost
condensation is the presence of long wavelength instabilities that evolve at a
very low speed $v_{\rm s}\simeq M/M_p$, and that will be triggered by classical
sources (and, it is natural to expect, by quantum fluctuations). An observer at
rest with respect to the preferred frame will see the growth of the instability
that, after a sufficiently long time, will show up as a measurable modification
of Newtonian gravity. An observer in motion, however, will not have the time to
see the development of the instability. As an analogy, we can think of a
overheated liquid, which, similarly to the ghost condensate, is in a metastable
state.  A small perturbation of the liquid -- such as the introduction of a
particle -- will lead to the generation of a bubble, which will then expand at
the speed of sound in the liquid. In this picture, the motion of a celestial
body in the ghost condensate is analogous to the motion of a particle quickly
traveling through this bubble chamber. Regions of modified gravity will be
nucleated where the source passes, but will then expand at the much smaller
velocity $v_{\rm s}\,$. An observer sitting on the particle will not have time
to see the growth of the bubbles. 

In the absence of observability of effects in tests of gravity, we expect the
signatures of the scenario to be associated to astrophysical or cosmological
effects. While in conventional gravity every source is associated to a
potential well, in this scenario every source will leave behind itself a
potential furrow, that will keep growing even when the source is far away.
Depending on the value of the parameter $M$, such unstable region will either
be still in its linear regime or will have evolved to a nonlinear stage. In the
latter case, it is possible to speculate about the existence of exotic compact
objects whose position would not be manifestly correlated with the present
distribution of matter. Even if the furrows are still in their linear regime,
they could manifest themselves as irregularities in the gravitational potential.
These irregularities could be felt by systems (such as the one formed by the
Earth and a satellite) that go across them. The possibility to detect them
through irregular motions of celestial bodies, or maybe also through lensing
effects, should be presumably studied statistically.

\vskip1pc
\noindent

\section{Acknowledgements}

It is a pleasure to thank Andy Albrecht, Paolo Creminelli, Nemanja Kaloper, Manoj Kaplinghat, Markus Luty, David Mattingly, Shinji Mukohyama, Keith Olive, Erich Poppitz, and Arkady Vainshtein for very fruitful discussions. The work of L.~S.\ was supported in part by the NSF Grant PHY-0332258.

\vspace{2cm}
{\bf Note Added:} After the completion of the present manuscript, we
became aware of the related work~\cite{dub}. The analysis and the
conclusions of~\cite{dub} are in agreement with the results presented
here.

\end{document}